\documentclass{article}
\usepackage{spconf,amsmath,graphicx}
\usepackage{cite}

\usepackage{CJK}

\title{Spike-Triggered Contextual Biasing for End-to-End Mandarin Speech Recognition}
\name{
\begin{tabular}{c}
\it Kaixun Huang$^{1}$, Ao Zhang$^{1}$, Binbin Zhang$^{2}$, Tianyi Xu$^{1}$, Xingchen Song$^{2}$, Lei Xie$^{1*}$\thanks{*Corresponding author.}
\end{tabular}
}
\address{$^1$Audio, Speech and Language Processing Group (ASLP@NPU), \\ Northwestern Polytechnical University, Xian, China \\
  $^2$WeNet Open Source Community}
\copyrightnotice{U.S. Government work not protected by U.S. copyright}

\begin{document}
%
\maketitle

\begin{abstract}
The attention-based deep contextual biasing method has been demonstrated to effectively improve the recognition performance of end-to-end automatic speech recognition (ASR) systems on given contextual phrases. However, unlike shallow fusion methods that directly bias the posterior of the ASR model, deep biasing methods implicitly integrate contextual information, making it challenging to control the degree of bias. In this study, we introduce a spike-triggered deep biasing method that simultaneously supports both explicit and implicit bias. Moreover, both bias approaches exhibit significant improvements and can be cascaded with shallow fusion methods for better results. Furthermore, we propose a context sampling enhancement strategy and improve the contextual phrase filtering algorithm. Experiments on the public WenetSpeech Mandarin biased-word dataset show a 32.0\% relative CER reduction compared to the baseline model, with an impressively 68.6\% relative CER reduction on contextual phrases.
\end{abstract}
\section{Introduction}
\label{sec:intro}

In recent years, thanks to the advancements in deep learning and the accumulation of massive data, end-to-end automatic speech recognition (ASR) methods, including connectionist temporal classification (CTC)~\cite{2006ctc, 2014ctc}, transducer~\cite{2019rnnt, 2021rnnt}, and attention-based encoder-decoder (AED)~\cite{2015aed, 2014aed, 2016aed}, have gained widespread application. However, in many real-world scenarios, audio data often contains phrases that are either scarce or completely absent from the training data. These phrases primarily consist of rare proper nouns and domain-specific terms. Although the overall error rate may not be significantly impacted by these phrases, the recognition accuracy of such phrases plays a crucial role in user experience and downstream tasks. For instance, in the scenario of voice assistants, the ASR system is required to accurately identify the names of individuals in the user's contact list whom they intend to call. Similarly, in the navigation scenario, the ASR system needs to accurately recognize the geographical names spoken by the user. Fine-tuning the model using relevant data can be a time-consuming process, rendering it impractical for rapidly changing and highly personalized scenarios. Therefore, integrating these phrases as contextual knowledge into ASR systems through \textit{contextual biasing} becomes increasingly important.

In prior studies, contextual biasing methods for end-to-end models have been roughly categorized into two groups: shallow fusion-based contextual bias~\cite{sf1, sf2, sf3, sf4, sf5}, such as weighted finite state transducer (WFST) based methods~\cite{sf2, sf3, sf4}, and neural attention-based contextual biasing (deep biasing)~\cite{2018db1, 2020db1, 2021db1, 2021db2, 2021db3, 2021db4, 2022db1, 2022db2, 2022db3, 2023db1, 2023db2, 2023db3, 2023db4, 2023db5}. Recently, due to their superior performance in enhancing bias phrases and their ease of integration into end-to-end models, \textit{deep biasing} methods have gained more popularity. Contextual Listen, Attend and Spell (CLAS)~\cite{2018db1} introduced for the first time a novel attention-based deep contextual biasing method, exploring contextual biasing approaches applicable to the Listen, Attend and Spell (LAS) model. Context-aware transformer transducer (CATT)~\cite{2021db1} integrates contextual information separately into the audio encoder and label encoder, providing a deep biasing technique applicable to the transducer model. Contextual phrase prediction network-based deep biasing method~\cite{2023db2}, by incorporating a bias loss to the audio encoder, enables an efficient contextual biasing approach across CTC, AED, and transducer models.

\begin{figure*}[thp]
\centering
\resizebox{0.7\textwidth}{!}{\includegraphics{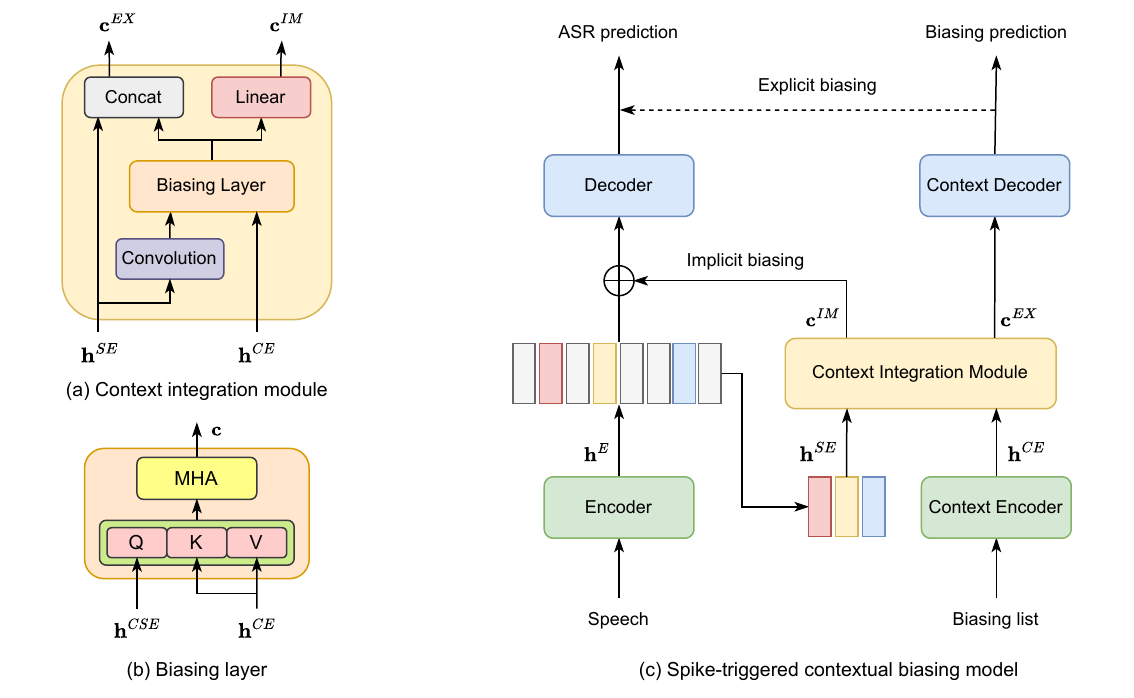}}

\caption{The proposed spike-triggered contextual biasing model}
\vspace{-0.2cm}
\end{figure*}

Most of the previous deep biasing methods share a common characteristic—the integration of contextual information is \textit{implicit} in nature. The biasing process of these methods can be summarized as follows. The context encoder encodes the biasing list to obtain contextual phrase embeddings, and then these embeddings are combined with the model's acoustic or label embeddings using an attention mechanism to obtain context-aware embeddings. Finally, these embeddings are concatenated or added to the model's acoustic or label embeddings for subsequent inference. In contrast to deep bias, shallow fusion methods often achieve biasing explicitly by modifying the posterior of the model's output during the decoding process. While implicit biasing methods may achieve better results, the unified structure makes it challenging to control the degree of biasing during the inference process. This limitation prevents setting higher bias levels for specific contextual phrases to achieve stronger biasing effects. Han et al.~\cite{2021db2} propose a collaborative decoding method on the continuous integrate-and-fire (CIF) based model for contextual biasing. It uses a context decoder to obtain the distribution of contextual outputs. Since the CIF module obtains token-level embeddings by integrating frame-level information, the context decoder has the same encoding granularity as the ASR decoder. This allows the context decoder to collaborate decoding with the ASR decoder to obtain the final hypothesis. As the collaborative decoding process operates on the final output distribution, it allows for \textit{explicit} control over the effect of contextual biasing. This method combines the advantages of shallow fusion and deep biasing, but is only applicable to CIF models, limiting its application to more popular ASR models.

To solve this problem, we propose a spike-triggered deep contextual bias method that supports both \textit{implicit} and \textit{explicit} bias in the AED model. To obtain the contextual biasing model, we freeze the parameters of the pre-trained ASR model and fine-tune the added contextual biasing module. We select emitting frames from the encoder output based on the CTC posterior and apply attention-based biasing only to these frames. During the inference process, when employing implicit biasing, context-aware embeddings are added to the emitting frames to achieve contextual biasing. When employing explicit bias, context-aware embeddings are used as input for the context decoder. The context decoder predicts the contextual phrases present in the utterance, and its posterior is used to bias the posterior for the ASR task. 

Furthermore, to further improve the effectiveness of contextual biasing, we propose a context sampling enhancement method, which replaces tokens in the contextual phrases and transcript with tokens that have similar pronunciations. We also optimize the contextual phrase filtering algorithm~\cite{2023db4} by computing it solely based on emitting frame posteriors and introducing a mismatch penalty score to prevent correct contextual phrases from being filtered out.

Experimental results on the open-source WenetSpeech Mandarin biased-word dataset~\cite{2023db3} demonstrate that our proposed method achieves a relative decrease of 32.0\% in the character error rate (CER) compared to the baseline model. Moreover, the CER for the contextual phrase (B-CER) portion decreases impressively by 68.6\%. Compared to the strong baseline model using the deep biasing method~\cite{2023db2}, our approach achieves a relative reduction of 6.2\% in B-CER. When combined further with the shallow fusion method based on WFST, B-CER decreases by 70.0\% compared to the baseline model. Our method exhibits good biasing effects on rare words that appear infrequently or have never been seen in the training set.

\section{Proposed Methods}
\label{sec:format}

\subsection{Model Structure}

Inspired by spike-triggered non-autoregressive ASR~\cite{spike}, we applied the spike-triggered approach for contextual biasing. Due to the nature of the CTC loss, there are numerous emitting spikes in the CTC posterior, with each spike corresponding to a decoded token. The spike-triggered method involves identifying these emitting spikes and using the encoder embedding of the frames corresponding to these spikes for subsequent computations. By employing the spike-triggered method, we can convert frame-level information into token-level information, which aids the subsequent biasing process and enables explicit biasing.

To achieve contextual biasing, we introduced three components into the standard AED model: the context encoder, context integration module, and context decoder, as shown in Figure 1(c).

The context encoder is responsible for encoding the variable-length contextual phrases from the biasing list into fixed-length context embeddings. It consists of a single layer of BLSTM. Each phrase in the tokenized list is tokenized using the same tokenizer as the base model. The resulting token sequence is then fed as input to the BLSTM, and the concatenated hidden states and cell states from the last time step of the forward and backward LSTM are passed through a linear layer to obtain the context embedding, denoted as $\mathbf{h}^{CE}$. Additionally, a contextual phrase consisting of blank tokens only is included in the biasing list, to enable the model to focus on it when there is no contextual phrase in the audio.

The context integration module is responsible for obtaining the context-aware embedding based on the encoder's emitting frame embedding and the context embedding. This module consists of a convolutional layer, a multi-head attention layer, and a linear layer, as shown in Figure 1(a).

First, the embedding of the emitting frames from the encoder, denoted as $\mathbf{h}^{SE}$, undergoes convolutional layers to obtain $\mathbf{h}^{CSE}$. The purpose of applying convolution is to incorporate information from the neighboring emitting frames into the embedding. This allows the model to consider more context information during subsequent attention operations, rather than relying solely on the information from the current emitting frame. Since the emitting frame embedding operates at the token level and the context embedding operates at the phrase level, applying convolution on the emitting frames helps reduce the granularity gap between the two.

In the multi-head attention layer, as shown in Figure 1(b), $\mathbf{h}^{CSE}$ serves as the query, and the context embedding $\mathbf{h}^{CE}$ serves as the key and value for calculating attention. This process extracts the context-aware embedding $c$ for each emitting frame, which can be described as follows:
\begin{equation}
\boldsymbol{\alpha}_{t}=\operatorname{Softmax}\left(\left(\mathbf{W}_{q} \mathbf{h}_{t}^{SE}\right)\left(\mathbf{W}_{k} \mathbf{h}^{CE}\right)^T / \operatorname{sqrt}\left(d\right)\right)~,
\end{equation}
\begin{equation}
\mathbf{c}_{t}=\boldsymbol{\alpha}_{t} \mathbf{W}_{v} \mathbf{h}^{CE}~,
\end{equation}
\vspace{0.1cm}
where the scaling factor $\operatorname{sqrt}\left(d\right)$ is for numerical stability. 

On one hand, the context-aware embedding undergoes further processing through a linear layer, resulting in $\mathbf{c}^{IM}$. Subsequently, $\mathbf{c}^{IM}$ is added back to its corresponding emitting frame embedding. On the other hand, $\mathbf{c}^{EX}$ is obtained by concatenating the context-aware embedding with the emitting frame embedding, and it serves as the input to the context decoder. This process can be described as follows:
\begin{equation}
\mathbf{c}^{IM}_t=\mathrm{Linear}(\mathbf{c}_t)~,
\end{equation}
\vspace{-0.5cm}
\begin{equation}
\mathbf{\tilde{h}}^{E}_t=\mathbf{h}^{E}_t+\mathbf{c}^{IM}_t,
\end{equation}
\vspace{-0.3cm}
\begin{equation}
\mathbf{c}^{EX}_t=[\mathbf{c}_t,\mathbf{h}^{SE}]~.
\end{equation}

The context decoder consists of a linear layer responsible for mapping the concatenated embedding $\mathbf{c}^{EX}$ to the posterior distribution over the vocabulary, predicting contextual phrases in the utterance. We keep the parts in the labels that are relevant to the contextual phrases unchanged, masking other tokens with the blank token. The bias loss for the contextual phrase prediction task is computed using the label smoothing loss function.

In addition to the AED loss, we incorporate auxiliary training using the CTC loss. Let $\lambda_1$ denote the weight assigned to the CTC loss and $\lambda_2$ denote the weight assigned to the bias loss. The joint loss function for the contextual biasing model can be expressed as follows:

\begin{equation}
\setlength\abovedisplayskip{6pt}
\setlength\belowdisplayskip{6pt}
L=\lambda_1 L_{ctc}+(1-\lambda_1) L_{att}+\lambda_2 L_{bias}~.
\end{equation}

During the process of training, we seek the optimal sequence of emitting frames on the posterior of the CTC based on the labels. We use the Viterbi algorithm to compute the optimal sequence of emitting frames and their corresponding posterior probabilities. During the inference phase, we adopt a similar approach to CTC greedy search. We consider frames where the maximum posterior probability token is not blank and differs from the maximum posterior probability token of the previous frame as emitting frames. Due to the fact that English ASR models often use byte pair encoding (BPE) tokens as modeling units, which are not segmented based on pronunciation, the emitting frames selected during inference are less accurate compared to Mandarin ASR models. This limitation affects the effectiveness of spike-triggered contextual biasing methods on English ASR models.

\subsection{Implicit/Explicit Contextual Bias}

When employing implicit contextual biasing for decoding, we retain the operation of adding $\mathbf{c}^{IM}$ back to the encoder emitting frame embedding while excluding the computation of the context decoder. The context-aware embeddings effectively convey the information of possible contextual phrases into the corresponding encoder emitting frame embedding, thus achieving implicit contextual biasing.

When employing explicit contextual biasing for decoding, we no longer add the context-aware embedding $\mathbf{c}^{IM}$ back to the encoder emitting frame embedding. Similar to the shallow-fusion bias method based on the WFST~\cite{sf3}, we use the bias list to construct a bias decoding graph. Figure 2 illustrates the bias decoding graph constructed using ``\begin{CJK*}{UTF8}{gbsn}语音识别\end{CJK*}" as the contextual phrase. In this graph, both $\phi$ arcs and $\rho$ arcs can match any token, with $\phi$ arcs not consuming the matched token while $\rho$ arcs consume the matched token. Dashed arcs are traversed only when no matching token is present.

We perform beam search with the posterior probabilities from the context decoder within the bias decoding graph. The path score is calculated as the sum of the context decoder posterior probability and the bias decoding graph score. To implement explicit biasing, we use decoding paths obtained through beam search, which exclusively consist of contextual phrases, to introduce bias to the posterior of the ASR task. Specifically, if a decoding path decodes a certain token at an emitting frame, a bias score is applied to that token in the posterior of the ASR task at this frame. Notably, bias scores are not cumulatively added when multiple decoding paths decode the same token in the same emitting frame. Compared to explicit biasing in CIF~\cite{2021db2}, our approach is more accurate due to the application of the bias decoding graph. This allows us to reduce the impact on non-biased parts and improve the effectiveness of explicit biasing.

\begin{figure}[t]
\centering
\label{figure2}
\resizebox{0.45\textwidth}{!}{\includegraphics{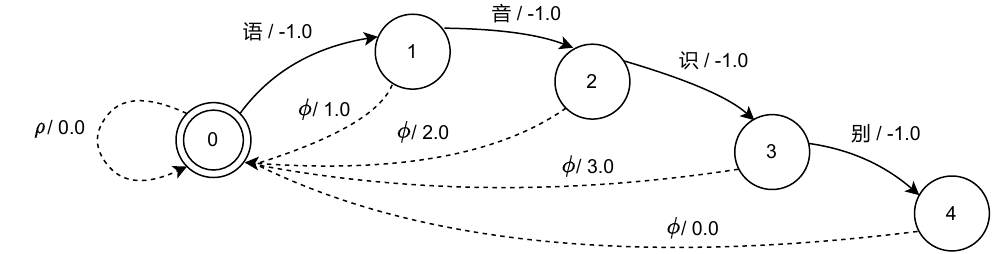}}

\caption{Example of bias decoding graph}
\vspace{-0.3cm}
\end{figure}

\vspace{-0.2cm}

\subsection{Contextual Phrase Filtering}

To mitigate the loss of recognition accuracy in the contextual biasing model when using a larger bias list, we employ a filtering mechanism on the list before inputting it into the context module. This process retains only those contextual phrases that are more likely to occur in the audio. We optimize the contextual phrase filtering algorithm based on Yang et al.'s work~\cite{2023db4}, by employing only the CTC posterior probabilities of emitting frames for more precise filtering.

To begin, we perform calculations using the bias list that only contains the blank token, resulting in obtaining unbiased CTC posterior probabilities. Moreover, we retain the audio embedding $\mathbf{h}^{E}$ throughout the process.

The contextual phrases are then filtered in two stages using
the unbiased CTC posterior. In the first stage, we calculate the posterior and the phrase score confidence (PSC). It computes the average of the maximum posterior values for each token present in the contextual phrases within a sliding window, without considering token order. In the second stage, we employ a dynamic programming algorithm to calculate the sequence order confidence (SOC). SOC considers the maximum average posterior values of the tokens within the sliding window, taking into account the token order. 

Building upon Yang et al.'s work~\cite{2023db4}, we introduced two optimizations. Firstly, we employ only posterior from emitting frames for filtering, thereby enhancing the efficiency of the filtering process. Additionally, we set a mismatch penalty score, denoted as $p$, equal to twice the confidence threshold, $q$. While calculating the PSC and SOC scores, the penalty score $p$ is permitted to represent the cost of insertion, deletion, and substitution errors. Specifically, even when the probability of a token in the posterior is very low, we can match that token with a cost of $p$ rather than using the exact posterior value of that token. Context phrases with PSC and SOC scores below the confidence threshold $q$ are filtered out. By introducing $p$, context phrases containing three or more tokens can be retained in the filtering process, even if one token has a very low posterior, as long as the probabilities of the other tokens are sufficiently high. This approach effectively improves the recall of the filtering algorithm.

After obtaining the filtered contextual phrases, we use the previously retained Encoder embedding $\mathbf{h}^{E}$ for subsequent biasing inference, resulting in more accurate biased recognition results.

\subsection{Context Sampling Enhancement Strategy}

To improve the effectiveness of contextual biasing, we employ a context sampling augmentation strategy during the training phase. We employ random sampling to extract contextual phrases from labels and enhance them by simultaneously replacing one token in the sampled contextual phrase and its corresponding token in the transcription with a token that shares the same pronunciation. Specifically, with a certain probability, we temporarily replace a token in the randomly sampled contextual phrase with a token that has a similar pronunciation. At the same time, we also temporarily replace the corresponding token in the transcript of the utterance. This requires the model to accurately recognize the replaced token through biasing, forcing it to learn to use contextual information to modify the recognition result.

\section{EXPERIMENTS}
\label{sec:pagestyle}

\subsection{Data}

Our experiments are conducted on the open-source Mandarin biased-word dataset~\cite{2023db3} based on the WenetSpeech corpus, consisting of 1000 hours of data. The test set is divided into three categories: personal names, place names, and organization names. The dataset uses the hanlp1 tool to identify named entities within the dataset. Only named entities that appear between 5 and 700 times in the training set are retained. This process is used to construct biasing lists for person names, location names, and organization names test sets.

We train the model using the entire 1000-hour training set and use the combined bias list (consisting of 298 contextual phrases) derived from merging the bias lists of the three test sets for evaluation. Additionally, to evaluate the model's ability to handle few-shot contextual phrases, we construct few-shot test sets using a subset of the WenetSpeech corpus. Specifically, we select named entities that appeared no more than 0, 3, 10, and 50 times in the training set and search for utterances in the WenetSpeech corpus that contain these named entities, resulting in the creation of 0-shot, 3-shot, 10-shot, and 50-shot test sets, respectively. Each few-shot test set consists of 400 utterances, with 100 contextual phrases.

\subsection{Experimental Setups}

\begin{table}[t]
\large
\renewcommand\arraystretch{1.7}
\caption{CER results on each test set. U-CER and B-CER represent the average results across all test sets.}
\vspace{0.1cm} 
\label{table1}
\resizebox{0.47\textwidth}{!}{
\begin{tabular}{lccccc}
\hline
\textbf{Model Type} & \textbf{Organization} & \textbf{Person} & \textbf{Place} & \textbf{U-CER} & \textbf{B-CER} \\ \hline
Baseline            & 10.42                 & 16.50           & 14.49          & 9.07           & 37.49          \\
CPPN                & 7.68                  & 10.67           & 10.42          & 9.09           & 12.54          \\
STCB-Implicit       & \textbf{7.32}                  & \textbf{10.49}           & \textbf{10.33}          & \textbf{9.02}           & \textbf{11.76}          \\
STCB-Explicit       & 7.99                  & 12.36           & 11.74          & 9.07           & 19.43          \\ \hline
\end{tabular}
}
\vspace{-0.2cm} 
\end{table}

\begin{table}[t]
\footnotesize
\centering
\renewcommand\arraystretch{1.7}
\caption{B-CER results on the few-shot test sets.}
\vspace{0.1cm} 
\label{table2}
\resizebox{0.45\textwidth}{!}{
\begin{tabular}{lcccc}
\hline
\textbf{Model Type} & \textbf{0-shot} & \textbf{3-shot} & \textbf{10-shot} & \textbf{50-shot} \\ \hline
Baseline            & 46.80           & 42.89           & 27.94            & 22.99            \\
STCB-Implicit       & 19.13           & 18.15           & 17.32            & 13.97            \\
STCB-Explicit       & 26.05           & 22.89           & 19.06            & 15.03            \\ \hline
\end{tabular}
}
\vspace{-0.2cm} 
\end{table}

The baseline model employs 12 conformer layers in the encoder and 6 transformer layers in the decoder, both with 256-dimensional inputs and 4 self-attention heads.

The spike-triggered contextual biasing model extends the baseline model by incorporating a context encoder, context integration module, and context decoder. The context encoder consists of 1 BLSTM layer and a linear layer. The context integration module includes a convolutional layer, 4-head attention layers, and another linear layer. The context decoder comprises a linear layer that projects the input dimension to the size of the vocabulary. The CTC loss weight $\lambda_1$ is set to 0.3, while the bias loss weight $\lambda_2$ is set to 0.2.

We use the contextual biasing model based on the contextual phrase prediction network~\cite{2023db2} as our deep biasing baseline. The configurations of its context encoder, biasing layer, and context decoder align with the corresponding modules in the spike-triggered contextual biasing model.

First, we conduct 60 rounds of training on the training set to obtain a baseline model by averaging the performance of the final 10 rounds. Then, we freeze the parameters of the baseline model and incorporate the context-aware modules. Fine-tune it for an additional 30 rounds to obtain the contextual biasing model. We use the attention rescore approach for decoding, which involves using the AED Decoder to score the results of the CTC beam search, combining the CTC and AED scores to obtain the final decoding results.

During the training phase, we randomly select three substrings with a length of 2 to 6 characters from the transcripts of each utterance. Additionally, we randomly select up to 5 contextual phrases from the biasing lists of other utterances as the biasing list for each utterance. We choose to use smaller biasing lists as the contextual phrase filtering algorithm is applied during inference. Using smaller biasing lists ensures consistency between training and inference. The probability of applying contextual phrase sampling augmentation is set to 0.1. The confidence threshold $q$ for the contextual phrase filtering algorithm is -6, and the penalty score $p$ is set to -12.

In addition to the CER, we also use the biased character error rate (B-CER) and unbiased character error rate (U-CER) to evaluate. U-CER measures the error rate for characters not present in the biasing list, while B-CER assesses the error rate for characters in the biasing list. Regarding insertion errors, if the inserted phrase is in the biasing list, it is counted towards B-CER and U-CER otherwise.

\subsection{Contextual ASR Accuracy}

In this section, we evaluate the performance of the baseline model and the contextual biasing model, referred to as STCB, comparing them with the Contextual phrase prediction network-based deep biasing method, referred to as CPPN~\cite{2023db2}. As shown in Table \ref{table1}, for each test set, both explicit and implicit contextual biasing based on spike-triggered achieve significant improvements in CER compared to the baseline model. The implicit biasing results in an average relative reduction of 32.0\% in CER and 68.6\% in B-CER compared to the baseline model. Moreover, the implicit biasing based on Spike triggered outperforms the CPPN method, which also employs implicit biasing.

It is worth noting that although contextual biasing is generally believed to have a negative impact on the non-contextual part of data, the U-CER of the contextual biasing model did not increase and even show a decrease in the test results. We speculate that this is due to the inclusion of fragments of contextual phrases from the biasing list in the test set. Since these fragments are not complete contextual phrases, they are not counted towards B-CER. However, they do benefit from contextual bias, leading to improved recognition accuracy and resulting in a decrease in U-CER.

\subsection{Few-shot Performance}

As shown in Table \ref{table2}, we conduct testing on few-shot datasets to evaluate the effectiveness of spike-triggered contextual biasing for rare words. For the implicit and explicit biasing methods, the relative decrease in B-CER on the 0-shot test set is 59.1\% and 44.3\% compared to the baseline model.  On the 10-shot test set, the relative reductions are 38.0\% and 31.8\%. The results indicate that our proposed method brings significant improvements, even for contextual phrases that have never appeared in the training set. Furthermore, the method performs best in enhancing the recognition of phrases that occur less frequently in the training set.

\subsection{Cascading with Shallow Fusion}

\begin{table}[t]
\Large
\renewcommand\arraystretch{1.7}
\caption{CER, U-CER, and B-CER results on the test set when cascaded with the shallow fusion method.}
\vspace{0.1cm} 
\label{table3}
\resizebox{0.48\textwidth}{!}{
\begin{tabular}{lccccc}
\hline
\textbf{Model Type} & \textbf{Organization} & \textbf{Person} & \textbf{Place} & \textbf{U-CER} & \textbf{B-CER} \\ \hline
SF                  & 7.78                  & 12.15           & 11.35          & 8.99           & 18.09          \\
STCB-Implicit + SF  & \textbf{6.81}                  & \textbf{10.19}           & \textbf{10.16}          & 8.77           & \textbf{11.23}          \\
STCB-Explicit + SF  & 7.09                  & 10.77           & 10.55          & 8.86           & 13.29          \\
STCB-All + SF       & 6.87                  & 10.20           & 10.19          & \textbf{8.75}           & 11.36          \\ \hline
\end{tabular}
}
\vspace{-0.2cm} 
\end{table}

Table \ref{table3} shows the performance of our proposed method when used in conjunction with the WFST-based shallow fusion bias method~\cite{sf1}. The shallow-fusion contextual biasing method, referred to as SF, also achieves good performance. After combining our method with the shallow fusion method, we observe further improvements, regardless of whether explicit or implicit biasing is used. For the explicit and implicit bias methods, the average relative reductions in B-CER on the test set are 64.6\% and 70.0\% compared to the baseline model. However, combining implicit biasing with explicit bias did not yield additional benefits. We believe this is because the information contained in explicit and implicit biases based on spike-triggered is similarity. The information provided by the shallow-fusion biasing method differs significantly from spike-triggered biasing, making the combined approach highly effective.

When cascading multiple methods to avoid excessive bias, it is necessary to add weight for each method. For explicit biasing and shallow fusion methods, we multiply the scores by this weight when adding scores during decoding. For implicit biasing, we multiply $\mathbf{c}^{IM}$ by this weight. In our first two experiments with cascading methods, we set the weight to 0.75. However, in the final experiment, it is set to 0.5.

\subsection{Contextual Phrase Filtering Performance}

\begin{table}[t]
\Large
\renewcommand\arraystretch{1.7}
\caption{CER, Recall, and Precision results of applying different contextual phrase filtering algorithms}
\vspace{0.1cm} 
\label{table4}
\resizebox{0.48\textwidth}{!}{
\begin{tabular}{lccccc}
\hline
\textbf{Model Type}  & \textbf{Organization} & \textbf{Person} & \textbf{Place} & \textbf{Recall} & \textbf{Precision} \\ \hline
STCB-Implicit        & 8.24                  & 11.40           & 11.57          & -               & -                  \\
+ vanilla filtering  & 7.48                  & 10.61           & 10.35          & 0.87            & 0.41               \\
+ improved filtering & 7.32                  & 10.49           & 10.33          & 0.91            & 0.45               \\ \hline
\end{tabular}
}
\vspace{-0.2cm} 
\end{table}

We compare the results of applying the vanilla contextual phrase filtering algorithm~\cite{2023db4} with our improved algorithm. From Table \ref{table4}, it is evident that using contextual phrase filtering is crucial when applying contextual biasing. Despite using a bias list size of only 298, without the use of the filtering algorithm, there is a noticeable degradation in CER. The improved contextual filtering algorithm demonstrates higher recall and precision for contextual phrases and achieves better CER results on all test sets.

\subsection{Ablation Study}

\begin{table}[t]
\Large
\renewcommand\arraystretch{1.5}
\caption{CER, U-CER, and B-CER results of ablation study}
\vspace{0.1cm} 
\label{table5}
\resizebox{0.48\textwidth}{!}{
\begin{tabular}{lccccc}
\hline
\textbf{Model Type} & \textbf{Organization} & \textbf{Person} & \textbf{Place} & \textbf{U-CER} & \textbf{B-CER} \\ \hline
STCB-Implicit       & 7.32                  & 10.49           & 10.33          & 9.02           & 11.76          \\
- convolution       & 7.54                  & 10.69           & 10.50          & 9.12           & 12.43          \\
- sampling enhancement   & 7.34                  & 10.58           & 10.48          & 8.96           & 12.61          \\
- bias loss         & 7.41                  & 10.54           & 10.47          & 9.08           & 12.08          \\ \hline
STCB-Explicit       & 7.99                  & 12.36           & 11.74          & 9.07           & 19.43          \\
- decoding graph    & 9.01                  & 13.69           & 12.63          & 8.99           & 26.01          \\ \hline
\end{tabular}
}
\vspace{-0.2cm} 
\end{table}

We conduct ablation experiments on the proposed method, and the results are shown in Table \ref{table5}. For the spike-triggered implicit biasing method, we remove the convolutional layer of the context integration module, context sampling enhancement, and bias loss separately. The results indicate that the absence of these processes weakens the effectiveness of contextual biasing. In contrast to the complete loss of bias ability when removing the bias loss in CPPN~\cite{2023db2}, the spike-triggered contextual biasing model still exhibits good bias ability even after removing the bias loss. This is likely because the spike-triggered mechanism transmits character-level alignment information to the model during training, allowing the model to learn how to perform contextual biasing even without relying on bias loss for training assistance. For the spike-triggered explicit bias method, we test the results of using the explicit biasing strategy from Han et al.~\cite{2021db2} on the same model. The results show that our method achieves better contextual biasing effects.

\section{CONCLUSIONS}
\label{sec:typestyle}

In this paper, we proposed a spike-triggered contextual biasing method that supports both explicit and implicit biasing on the AED model. We also introduced a context sampling enhancement strategy during the training phase and improved the contextual phrase filtering algorithm. Compared to previous deep bias methods, our approach achieves better bias effects in the Mandarin dataset. Additionally, our method can be cascaded with shallow fusion methods, leading to further improvements in contextual phrases. Due to the relatively low accuracy of emitting frame selection during inference in ASR models modeled with BPE, the effectiveness of our method on such models is not significant. Hence, we will further explore more robust methods for selecting emitting frames during inference in our future investigations.

\bibliographystyle{IEEEbib}
\bibliography{strings,refs}

\end{document}